\documentclass[5p]{elsarticle}

 \usepackage{graphicx}

\usepackage{amssymb}

\usepackage[nodots]{numcompress}

 \usepackage{lineno}

\journal{Nuclear Physics B}

\begin{document}

\begin{frontmatter}


\title{TeV sky versus AUGER one: are UHECR also radioactive, heavy galactic nuclei?
   }

\author{Daniele Fargion}
\ead{daniele.fargion@roma1.infn.it}
\address{Departments of Physics, Universit\`{a} di Roma 1, Sapienza
and INFN --Pl. A. Moro 2, 00185, Rome, Italy.}

\begin{abstract}
  UHECR (Ultra High Cosmic Rays)  made by He-like lightest nuclei might  solve the AUGER extragalactic clustering along Cen A: He UHECR cannot arrive from Virgo  because the light nuclei fragility and opacity above few Mpc;  UHECR signals are clustering along Cen-A\ spreading as observed by horizontal galactic arms magnetic fields, along random vertical angles  \cite{Fargion2008}. Cen A events by He-like spread along a width angle as large as the observed clustered one; such a light nuclei coexist with the  Auger heavy nuclei and with the Hires nucleon composition \ref{fig4}. UHECR He, being fragile should partially fragment in secondaries  at  tens EeV  multiplet (D,$He^{3}$,p) almost as it occurs in the very recent UHECR multiplet at $20$ EeV  along Cen A UHECR clustering  \cite{Fargion2011}. Their narrow crowding  within a tiny (Cen A-centric disk) observation area (below $10^{-2}$ of the whole AUGER sky) aligned with Cen A may occur by a very low probability, below $3\cdot 10^{-5}$, see Fig. \ref{fig0}, \ref{fig1},\ref{fig2}, \cite{Fargion2011}. Remaining UHECR spread group  show correlations with other gamma  (MeV-$Al^{26}$ radioactive) maps, mainly  due to galactic SNR sources as Vela pulsar, the brightest, nearest GeV source \cite{Fargion09b}. Other nearest galactic gamma sources (as partially Crab and Galactic Center core)  show links with UHECR via TeV correlated maps, see Fig.\ref{fig4-5}. We speculate here that UHECR are also heavy radioactive galactic nuclei  as $Ni^{56}$,  $Ni^{57}$  and $Co^{57}$,$Co^{60}$ widely bent (tens degree up to $\geq 100^{o}$)  by galactic fields.  UHECR radioactivity (in $\beta$ and $\gamma$ channels) and decay in flight at hundreds keV  is boosted (by huge Lorentz factor $\Gamma_{Ni}\simeq 10^{9}- 10^{8}$)  leading to TeVs gamma correlated sky anisotropy, see Fig.\ref{fig4-5}. Tau neutrinos secondaries and their tau airshowers at horizons may rise (from hundreds TeVs-PeVs) in future ICECUBE,ANTARES \cite{antares}, double bang events \cite{Learned}, or in  ARGO and ASHRA \cite{Aita11} horizons or in AUGER or TA airshowers at PeVs-EeV energies, showing  the expected Neutrino astronomy link to UHECR decay in flight.
   \end{abstract}


\end{frontmatter}



\section{UHECR composition and maps}
UHECR astronomy is becoming a reality, suffering however by directionality  smearing of magnetic field along the UHECR arrival directions \cite{Auger-Nov07}.
   The UHECR compositions are leading to different bending and maps, different fragment and secondary gamma, neutrino spectra; these different nature are making UHECR nucleon origination  well directed (but in a GZK bounded , tens Mpc distances, Universe) or even much local and smeared (a few Mpc) for our lightest UHECR nuclei. If UHECR are mostly heavy as Fe or Ni,Co, as argued in present paper (see also \cite{Fargion2011c}) then UHECR astronomy will be  so much bent and polluted to be greatly smeared  and hardly correlated to their sources. However very local galactic sources maybe recognized. On the other side lightest nuclei UHECR astronomy may lead to a parasite  astronomy (i.e an astronomy associated due to secondaries fragments of UHECR as pions and consequent gamma, charged leptons and neutrinos) observable as a MeV-TeV gamma, UHE neutrino, and UHECR lightest nuclei fragment; these traces might be partially source of  radio, X  tails. This is the case of unique extragalactic nearest AGN, Cen A.  Extragalactic heavy nuclei may also be traced by nuclei fragments but at much lower energy. However we are considering also galactic UHECR heavy nuclei whose radioactivity is source of parasite gamma TeVs anisotropy in the sky.   Indeed we  address here on the UHECR anisotropy nature able  to  correlate UHECR maps and composition. Inspired by MeV map correlation we try to find a reading key also for TeV-UHECR correlations. Heavy radioactive nuclei  offer a tuned solution for UHECR excluding Cen A area.
\subsection{UHECR from Cen A by He and its fragments}
  Extragalactic UHECR formed (mostly) by  lightest nuclei may explain a partial clustering of events, as the one around CenA \cite{Auger10},\cite{Fargion2011} as well as the puzzling UHECR absence  around Virgo \cite{Auger-Nov07}. Light nuclei are fragile and fly few Mpc before being    halted by photo-disruption \cite{Fargion2008}.   Of course a more heavy UHECR nature may lead to a complete  or partial confinement explaining in alternative the Virgo absence. A very tuned model may suggest that UHECR from Virgo are sitting along the apparent Cen A clustering of events \cite{Semikoz10}. We considered this a very ad hoc model \cite{Fargion2011} because of the tuned coincidence of a wide area (Virgo) bent and focalized in the narrow clustering Cen A area. The light nuclei fragments $He + \gamma \rightarrow D+D,D+\gamma \rightarrow p+n+ \gamma  , He + \gamma \rightarrow He^{3}+n, He + \gamma \rightarrow T +p $  may nevertheless trace the same UHECR maps by a secondary  clustering at half or even fourth of the UHECR primary  energy \cite . At lower energy (at ten EeV or below) the huge smeared cosmic ray isotropy and homogeneity may hide these tiny inhomogeneity traces  \cite{Fargion2010},\cite{Fargion2011}. About the extragalactic hypothesis let us remind that gamma UHECR secondaries rays  are partially absorbed by microwave and infrared background making once again a very local limited UHECR-gamma astronomy. Muon neutrinos $\nu_{\mu}$, the most penetrating and easy to detect on Earth, are unfortunately deeply polluted by atmospheric (homogeneous) component (as smeared and as isotropic as their parent charged CR nucleons and nuclei ). This atmospheric neutrino isotropy and homogeneity  (made by primary CR bent and smeared by galactic magnetic fields)
   is well probed by last TeV muon neutrino smooth ICECUBE map. Tau neutrinos on the contrary, the last neutral lepton discovered, are almost absent in atmospheric secondaries (about five order of magnitude suppressed, because they are not produced easely by charmed process). At a few GeV energies $\tau$ cannot be born because energy thresholds. Only at ten GeV  atmospheric windows, $\nu_{\mu}\rightarrow \nu_{\tau} $ neutrino oscillation   may  arise because the Earth size is large  enough to allow a complete neutrino $\nu_{\mu}\rightarrow \nu_{\tau} $  conversion; at tens TeV-PeV up to EeV  $\nu_{\tau} $ atmospheric neutrinos cannot convert within short Earth size (from $\nu_{\mu}\rightarrow$ $\nu_{\tau}$) because they are unable to oscillate at such high energies; therefore $\nu_{\tau}$ neutrino might be a clean signal of  UHECR-neutrino associated astronomy\cite{FarTau}. Their tau birth in ice or sea  may shine as a double bangs (disentangled above PeV)  \cite{Learned} or in a \emph{mini twin bangs} observable in Deep Core  or PINGU detector \cite{Fargion2011c}. In addition high energy UHE $\nu_{\tau}$ and its ${\tau}$, born tangent to the  Earth or mountain, while escaping  in air  may lead, by decay in flight, to  loud, amplified  and well detectable directional tau-airshower at horizons \cite{Fargion1999},\cite{FarTau}. Both in atmosphere fluorescence tracks or by Cherenkov blazing \cite{Aita11} \cite{FarTau}, or by partial skimming ground detectors \cite{FarTau} \cite{Feng02}. Tau astronomy versus UHECR are going to reveal most violent  sky as the most deepest probe. This tau airshowers or Earth skimming neutrino \cite{Feng02} were considered since more than a decade and are going to be observed in AUGER or TA in a few years \cite{Fargion1999},\cite{FarTau},\cite{Auger-01},\cite{Feng02},\cite{Auger07}.
\subsection{Overlapping Gamma and UHECR maps}
     We  found that Cen A (the most active and nearby extragalactic AGN) is apparently the most shining  UHECR source whose clustering (almost a quarter of the event) along a narrow solid angle  (whose opening angular size is  $\simeq 17^{o}$) \cite{Auger10}, is convincing and in agreement with lightest nuclei \cite{Fargion2008},\cite{Fargion09a}, \cite{Fargion09b}, \cite{Fargion2009}.
     However the main question is related to remaining majority of events. Where do they come from?
          In recent maps of UHECR we noted a first hint of  Vela see Fig.\ref{fig4-5}, where the brightest and one of the nearest gamma source, is associated to an unique UHECR triplet nearby the pulsar \cite{Fargion09b}. The correlation is also based on MeV  Comptel map  and by other  coincidence \cite{Fargion09b}, see Fig . The needed UHECR  bending from near Vela ($815$ yc) is calling for a very heavy nuclei (or light nuclei with large magnetic field).
               Remaining UHECR events might be also mostly   heavier nuclei more bent and smeared by galactic fields.
Let us remind that UHECR events initially consistent with  GZK volumes  \cite{Auger-Nov07}, today seem to be not much correlated with expected Super Galactic Plane \cite{Auger10}. Moreover slant depth data of UHECR from AUGER airshower shape do not favor the proton but point to  a nuclei. Therefore in the present paper we suggest UHECR as made by nuclei, light from Cen A and heavy mostly from our galaxy. However it maybe worth to briefly remind the long (two decades at least) history of the UHECR understanding somehow linked to a century old (Victor Hess first balloon detection) cosmic ray puzzle.

\section{A two decades of random UHECR puzzles}
Let us remind that Cosmic Black Body Radiation had imposed since $1966$ a cut, GZK cut-off \cite{Greisen:1966jv}, of highest energy cosmic ray propagation, implying a very limited cosmic Volume (ten or few tens Mpc) for highest UHECR  nucleon events. Because of the UHECR rigidity one had finally to expect to track easily UHECR directionality back toward the sources into a new Cosmic Rays Astronomy.
Indeed in last two decades, namely since 1991-1995 the rise of an  \emph{unique, apparent} UHECR event at $3\cdot {10^{20}}$ eV, by Fly's Eye,  has opened the wondering of its origination: no nearby (within GZK cut off) source have been correlated. Many exotic decaying topological relics have been advocated. But no clear candidature survived. Incidentally it should be noted that even after two decades  and after a huge increase of aperture observation ( nearly two order of magnitude in area-time by the records in  AGASA, HIRES, AUGER, TA) no equal or larger energy event as the first $3 {10^{20}}$ eV has been rediscovered. This remarkable silence  makes a shadows on that exceptional initialstarting UHECR event. Anyway  to face the uncorrelated UHECR at $3\cdot {10^{20}}$ eV event, and later on the events by AGASA, the earliest evidences (by SuperKamiokande) that neutrinos have a non zero mass had opened  \cite{Fargion1997} the possibility of  an UHECR Neutrino-mass connection:  UHE ZeV neutrino (originated by AGN or GRB at far redshift) could be the transparent  courier of any far source (beyond GZK radius); such ZeV $\nu$  may hit and scatter on  a local relic anti-neutrino dark halo (or viceversa over neutrino), spread  as a dark cloud at a few Mpc around our galaxy. Its resonant$\nu$$\bar{\nu}$ Z boson (or WW pair channel) production  \cite{Fargion1997},\cite{Weiler},  is source of a secondary nucleon later on observable at Earth as a UHECR. The rare and uncorrelated $3\cdot {10^{20}}$ eV  event could be associated to a very far AGN (born by a Seyfert galaxy MCG 8-11-11) whose distance is well above GZK cut off (it is located at a redshift of z=0.0205, nearly 80Mpc, above 10 Mpc GZK cut). The later search by AGASA seemed to confirm the Fly's Eye events above $ {10^{20}}$eV  and the puzzling absence of nearby expected GZK anisotropy or correlation.  On $1999-2000$ t he absence of the UHECR GZK cut-off was generally accepted. In different occasion HIRES data offered a possible  UHECR connections with far BL Lac \cite{Gorbunov}, giving hopes and argument to our Z-resonant (Z-burst) model \cite{Fargion1997}.
However the evidence for a low neutrino mass and the more recent results obtained by Hires, with increased effective area, and AUGER (2001-2007) allowed to claim the evidences for a GZK suppression in UHECR spectra.  But in addition AUGER has shown an anisotropic clustering, on early 2007 events located apparently along the Super Galactic Plane, a place well consistent with GZK expectation \cite{Auger-Nov07}. Therefore if that was the end, than the UHE neutrino scattering on relic dark matter neutrino ones \cite{Fargion1997}, became obsolete. But, again this fast AUGER conclusions seem today unnecessary, and UHECR puzzle remained (out of our present interpretation) mostly unsolved. The very  exceptional  blazar $3C454.3$ flare on $2nd December$ $2009$, the few AGN connection of UHECR located much far from a GZK distances may suggest us, surprisingly, to reconsider such exceptional model \cite{Fargion1997}. Possibly connecting lowest neutrino   particle ($\simeq 0.15$ eV) mass with highest UHE ($\simeq 30$ $ {10^{21}}$eV) neutrino energies.
 Indeed here we review the most recent  UHECR maps \cite{Auger10} over different  ones  and we comment some  feature, noting as mentioned above, some possible partial galactic component \cite{Fargion2010},\cite{Fargion09b}. The consequences of the UHECR composition and source distances should reflect, by interaction with relic photons, to UHE  secondaries signals made by photo-pion production or photo-nuclear disruption  as the well known GZK \cite{Greisen:1966jv} or cosmo-genic neutrinos. The proton UHECR provide EeV neutrinos (muons and electron) whose flavor oscillation lead to tau neutrinos to be soon detectable \cite{FarTau} \cite{Auger08} by upward tau air-showers;  the UHECR lightest nuclei model provide only lower energy, tens PeV, neutrinos detectable in a very peculiar way by AUGER fluorescence telescopes or in ARGO array by horizontal $\tau$ air-showers  , or by Icecube $km^3$ neutrino telescopes \cite{FarTau},\cite{Fargion2009},\cite{Fargion09a}  \cite{Fargion09b} either by double bang\cite{Learned}, or by long muon at few PeV energy. Let us recall that ZeV UHE neutrinos in Z-Shower model are also possible source of horizontal Tau air-showers of maximal size and energy \cite{FarTau},\cite{Fargion2010}.
\section{The Lorentz UHECR bending for $He^{4}$ and  $Ni^{57}$ }
Cosmic Rays are blurred by magnetic fields. Also UHECR suffer of a Lorentz force deviation. This smearing maybe source of UHECR features. Mostly along Cen A.
 There are at least three mechanisms for magnetic deflection along the galactic plane, a sort of galactic spectroscopy of UHECR. The magnetic  bending by extra-galactic fields are  in general negligible respect galactic ones.
 A late nearby (almost local) bending by a nearest coherent galactic arm field, and a random one by turbulence and a random along the whole plane inside different arms:\\
(1)
The coherent Lorentz angle bending $\delta_{Coh} $ of a proton (or nuclei) UHECR (above GZK \cite{Greisen:1966jv}) within a galactic magnetic field  in a final nearby coherent length  of $l_c = 1\cdot kpc$ is $ \delta_{Coh-p} .\simeq{2.3^\circ}\cdot \frac{Z}{Z_{H}} \cdot (\frac{6\cdot10^{19}eV}{E_{CR}})(\frac{B}{3\cdot \mu G}){\frac{l_c}{kpc}}$.\\
(2) The random bending by random turbulent magnetic fields, whose coherent sizes (tens parsecs) are short and whose final deflection angle is  smaller than others and they are here ignored.\\
(3) The ordered multiple UHECR bending along the galactic plane across and by alternate arm magnetic field directions whose final random deflection angle is remarkable and discussed below.\\

The bending angle value is  quite different for a heavy nuclei as an UHECR from Vela whose distance is only $0.29$ kpc:
 $$\delta_{Coh-Ni} \simeq
{18,7^\circ}\cdot \frac{Z}{Z_{Ni^{28}}} \cdot (\frac{6\cdot10^{19}eV}{E_{CR}})(\frac{B}{3\cdot \mu G})({\frac{l_c}{0.29 kpc})}$$
  Note that this spread is  able to explain the nearby Vela TeV anisotropy (because radioactive emission in flight) area around its correlated UHECR triplet.
There is an extreme possibility: that Crab pulsar ar few kpc is feeding the TeV anisotropy connecting with a gate its centered disk to a wider extended region where some UHECR are clustering. From far Crab distances the galactic bending is:  $$\delta_{Coh-Ni} \simeq
{129^\circ}\cdot \frac{Z}{Z_{Ni^{28}}} \cdot (\frac{6\cdot10^{19}eV}{E_{CR}})(\frac{B}{3\cdot \mu G})({\frac{l_c}{2 kpc})}$$
  Note that such a spread is able to explain the localized TeV anisotropy born in Crab (2 kpc)apparently  extending  around  area near Orion, where also spread UHECR events seem clustered.
Such heavy iron-like  (Ni,Co) UHECR , because of the bug charge and large angle bending, are mostly bounded inside a Galaxy, as well as in Virgo cluster, possibly explaining the UHECR  absence in that direction. We shall not discuss  here rare (and in our view un-probable) large angle coherent deflection from Virgo tuned and overlayed into Cen A direction as some authors suggested  \cite{Semikoz10}.

 The incoherent random angle bending (2) along the galactic plane and arms, $\delta_{rm} $, while crossing along the whole Galactic disk $ L\simeq{20 kpc}$  in different (alternating) spiral arm fields   and within a characteristic  coherent length  $ l_c \simeq{2 kpc}$ for He nuclei is $$\delta_{rm-He} \simeq{16^\circ}\cdot \frac{Z}{Z_{He^2}} \cdot (\frac{6\cdot10^{19}eV}{E_{CR}})(\frac{B}{3\cdot \mu G})\sqrt{\frac{L}{20 kpc}} \sqrt{\frac{l_c}{2 kpc}}$$ The heavier  (but still light nuclei) bounded from Virgo might be also Li and Be:
 $$\delta_{rm-Be} \simeq{32^\circ}\cdot \frac{Z}{Z_{Be^4}} \cdot (\frac{6\cdot10^{19}eV}{E_{CR}})(\frac{B}{3\cdot \mu G})\sqrt{\frac{L}{20 kpc}}
\sqrt{\frac{l_c}{2 kpc}}$$.  It should be noted that the present anisotropy above GZK \cite{Greisen:1966jv} energy $5.5 \cdot 10^{19} eV$ (if extragalactic)  might leave a tail of signals: indeed the photo disruption of He into deuterium, Tritium, $He^3$ and protons (and unstable neutrons), rising as clustered events at half or a fourth (for the last most stable proton fragment) of the energy:\emph{ protons being with a fourth an energy but half a charge He parent may form a tail  smeared around Cen-A at twice larger angle} \cite{Fargion2011}. We suggested  to look for correlated tails of events, possibly in  strings at low $\simeq 1.5-3 \cdot 10^{19} eV$ along the Cen A train of events. \emph{It should be noticed that Deuterium fragments have half energy and mass of Helium: Therefore D and He spot are bent at same way and overlap into UHECR circle clusters}\cite{Fargion2011}.  Deuterium are even more bounded in a very local Universe because their fragility (explaining Virgo absence). In conclusion He like UHECR maybe bent by a characteristic angle as large as $\delta_{rm-He}  \simeq 16^\circ$; its expected lower energy Deuterium tails at half energy ($30-25 EeV$)  also at ($\delta_{rm-p}  \simeq 16^\circ$); protons last traces at a quarter of the UHECR energy, around twenty EeV multiplet, will be spread within ($\delta_{rm-p}  \simeq 32^\circ$): within the observed CenA UHECR multiplet  in figures.
 \subsection{Twin UHECR multiplet at 20 EeV pointing Cen A}
\begin{figure}[!t]
  \vspace{5mm}
  \centering
  \includegraphics[width=3.2 in]{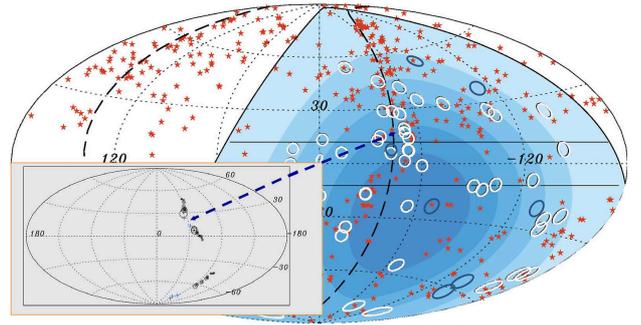}
  \caption{The last 2010 UHECR event map \cite{Auger10}  and the multiplet event map \cite{Auger11}: two of the three train are clustering toward Cen A, whose pointing sources are marked by crosses as shown by the arrow \cite{Fargion2011} }
  \label{fig0}
 \end{figure}

\begin{figure}[!t]
  \vspace{5mm}
  \centering
  \includegraphics[width=3.1 in]{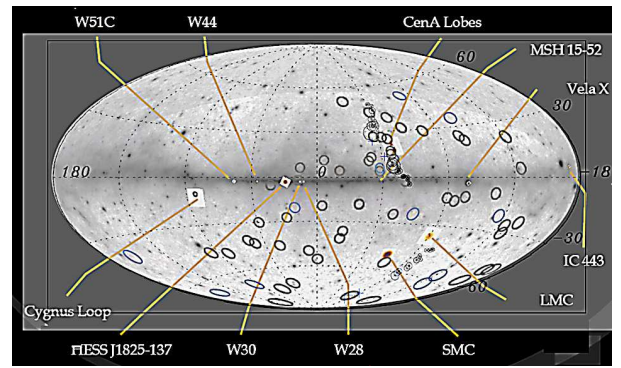}
  \caption{The last 2010 UHECR event map \cite{Auger10} by AUGER and the overlap multiplet clustering toward Cen A \cite{Auger11}, inside the last Fermi 2011 gamma map and labels, whose UHECR twin multiplet expected sources are within a tiny disk area (of  radius below $7.5^{o}$)}
  \label{fig1}
 \end{figure}

\begin{figure}[!t]
  \vspace{5mm}
  \centering
  \includegraphics[width=3.1 in]{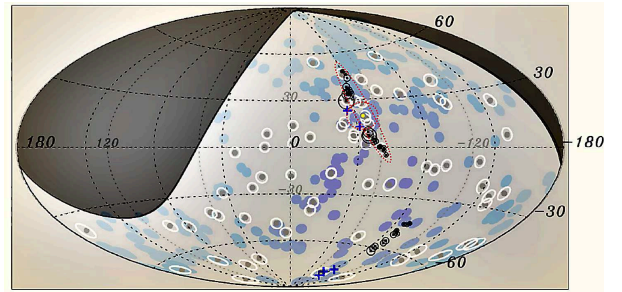}
  \caption{The AUGER 2010 UHECR event map  \cite{Auger10} and two of the three multiplet clustering toward Cen A \cite{Auger11}; their sources as shown by dotted curve are  within a tiny disk area (at radius of $7.5^{o}$); the dotted ellipsoid area of the UHECR and multiplet clustering is also extremely correlated, aligned and small. }
  \label{fig2}
 \end{figure}

\begin{figure}[!t]
  \vspace{5mm}
  \centering
  \includegraphics[width=2.7 in]{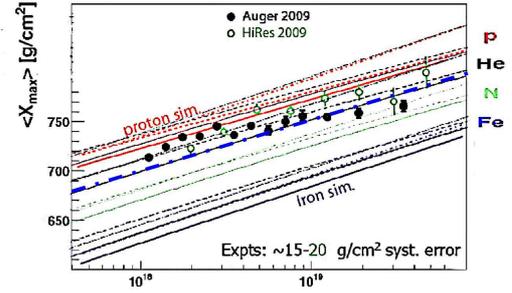}
  \caption{One recent UHECR AUGER slant depth and composition derived by air shower feature; note the best fit of He on most of the highest UHECR events combining both Hires and AUGER results}
  \label{fig3}
 \end{figure}

\begin{figure}[!t]
  \vspace{5mm}
  \centering
  \includegraphics[width=2.5 in]{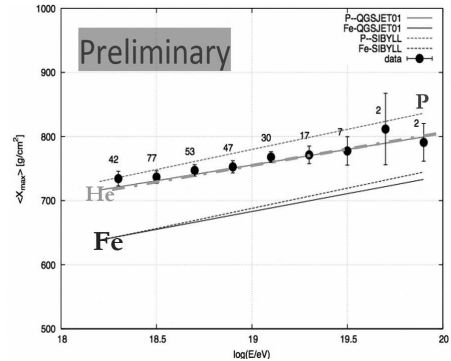}
  \caption{One of  last UHECR Telescope Array Composition derived by air shower slant depth shown on 2011; note the best fit of He on most of the highest UHECR events combining both Hires and AUGER results}
  \label{fig4}
 \end{figure}

 The very  recent multiplet clustering published just few weeks ago by AUGER at twenty EeV  contains just three and apparently isolated train of events  with  (for the AUGER collaboration) no statistical meaning. \cite{Auger11}. Indeed apparently they are pointing to unknown sources (See Fig. \ref{fig0}). However the crowding of the two train multiplet tail centers inside a very narrow disk area focalized about the rarest Cen A UHECR source is remarkable (See Fig. \ref{fig1}).  If UHECR are  made by proton (as some AUGER author believe) they will not naturally explain such a tail structure because these events do not cluster more than a few degree, contrary to observed UHECR and associated multiplet (See Fig. \ref{fig2}). Also heavy nuclei whose smearing is much larger and whose eventual nucleon fragments ($A\rightarrow (A-1)$) should lead to parasite tail that greatly differ  in mass and energy and bending angle with the observed AUGER one. If heavy UHECR are around and bent only a galactic smeared component may be somehow discovered (for the composition see Fig.\ref{fig3})(see Fig.\ref{fig4}),( for gamma association map see Fig.\ref{fig4}). Our He-like UHECR do fit the AUGER and the HIRES composition traces (see Fig.\ref{fig3})(see Fig.\ref{fig4}). The He secondaries are splitting in two (or a fourth) energy fragments along Cen A tail (see Fig.\ref{fig2})  whose presence has  being foreseen and published many times  in last years\cite{Fargion2011}. Indeed the  dotted circle around Cen A containing the two (of three) multiplet  (see Fig.\ref{fig2}) has a radius as small as $10^{o}$,  it extend in an area that is as smaller as  $314$ square degrees, below or near $1\% $ of the observation AUGER sky (see Fig.\ref{fig0}). The probability that two among three sources fall inside this small area is offered by the binomial distribution. $$ P (3,2) = \frac{3!}{2!} \cdot (10^{-2})^{2} \cdot \frac{99}{100}\simeq 3 \cdot 10^{-4}$$. Moreover the same twin tail of the events are aligned almost exactly $\pm 0.1 $ rad along UHECR train of events toward Cen A (see Fig.\ref{fig2}). Therefore the UHECR  multiplet alignment  at twenty EeV has an a priori  probability as low as $ P (3,2) \simeq 3 \cdot 10^{-5}$ to follow the foreseen signature\cite{Fargion2011}.

\section{TeV Gamma and UHECR  nuclei connection? }
\begin{figure}[!t]
  \vspace{5mm}
  \centering
  \includegraphics[width=3.2 in]{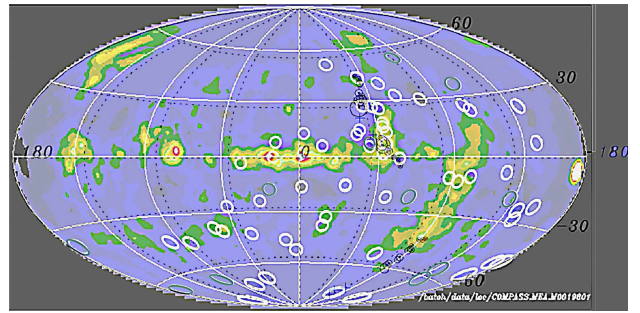}
  \caption{The last 2010 UHECR event map by AUGER and the Multiplet clustering toward Cen A overlap the MeV Comptel gamma map; note the apparent clustering of UHECR along the Vela, Magellanic stream , Cen A and other galactic regions. These gamma area may contains additional clustering in future records probing a  galactic nature of a fraction of UHECR.}
  \label{fig5}
 \end{figure}

As we mentioned Cen A may explain nearly a quarter of the UHECR events; but where are the other come from?
What is the nature and origin of all the other UHECR events? Composition favor mostly  heavy nuclei and as we shall see in figure below by local sources.
       In recent maps of UHECR we noted first hint of galactic source   rising as an UHECR triplet \cite{Fargion09b}.  Also the hint by $Al^{26}$ gamma map traced by Comptel somehow overlapping with UHECR events at 1-3 MeV favors a role of UHECR radioactive elements (as $Al^{26}$), see Fig. \ref{fig5}.   The most prompt ones are the  $Ni^{56}$, $Ni^{57}$ (and $Co^{56}$, $Co^{60}$ ) made by Supernova (and possibly by their collimated GRB micro-jet components see \cite{Fargion1998}) ejecta in our own galaxy.  Indeed in all SN Ia models, the decay chain  $Ni^{56}\rightarrow Co^{56} \rightarrow Fe^{56}$provides the primary source of energy that powers the supernova
optical display. The $Ni^{56}$ decays by electron capture and the daughter $Co^{56}$  emits gamma rays by the nuclear de-excitation
process; the two characteristic gamma lines are  respectively at $E_{\gamma} =$ 158 keV and $E_{\gamma}=$ 812 keV. Their half lifetime are spread from $35.6$ h for  $Ni^{57}$ and $6.07$ d. for   $Ni^{56}$. However there are also more unstable radioactive rates as  for $Ni^{55}$ nuclei whose half life is just $0.212$ s or $Ni^{67}$ whose decay is $21$ s. Therefore we may have an apparent boosted  ($\Gamma_{Ni^{56}} \simeq 10^{9}$) life time spread from $2.12 \cdot 10^{8}$ s or $6.7$ years (for $Ni^{55}$) up to nearly  $670$ years (for $Ni^{67}$) or $4$ million years for  $Ni^{57}$. This consequent wide range of lifetimes guarantees a long life activity on the UHECR radioactive traces. However the most bright are the fastest  decaying ones. The arrival tracks of these UHECR radioactive heavy nuclei may be  widely bent, as shown below, by galactic magnetic fields. Among the excited nuclei to mention  for the UHECR-TeV connection is $Co_{m}^{60}$  whose half life is $10.1$ min and whose decay gamma line is at $59$ keV. At a boosted nominal Lorentz factor $\Gamma_{Co^{60}}= 10^{9}$ we obtain $E_{\gamma}\simeq 59 $ TeV ; let us remind that a gamma air-shower exhibit a smaller muon trace and it simulates a ($10\%$) hadronic shower ( $E_{gamma-hadron}\simeq 6 $ TeV) nearly corresponding to observed anisotropy. The decay boosted lifetime is $19000$ years or 6kpc distance. Therefore $Co_{m}^{60}$ energy decay traces, lifetime and spectra fit well within  present UHECR-TeV connection for nearby galactic sources as Vela and (probable) Crab. Other radioactive beta decay, usually at higher energy may also shine at hundred or tens TeV or below by inverse Compton and synchrotron radiation. Therefore their UHECR bent parental nuclei may shine also in TeV Cosmic ray signals (see Fig. \ref{fig4-5})


\section{TeVs along Cygnus and Crab: Boosted and bent $Ni^{57}$-$Co^{60}$ decaying nuclei?}
The very rich UHECR map of AUGER and HIRES in celestial coordinate overlapped on TeV anisotropy  (North sky by ARGO- South sky by ICECUBE atmospheric muons) is displayed in next figure \ref{fig4-5}.
 The figure shows a clear area around Crab that is somehow extending in a wide anisotropy area where few UHECR took place. We introduced some arrows to label these linked area; we also added arrows to
 remind the Cen A unique clustering as well as the Vela and the Cygnus galactic TeV sources, somehow in connection with UHECR.

 \begin{figure}[!t]
  \vspace{5mm}
  \centering
  \includegraphics[width=3.6 in]{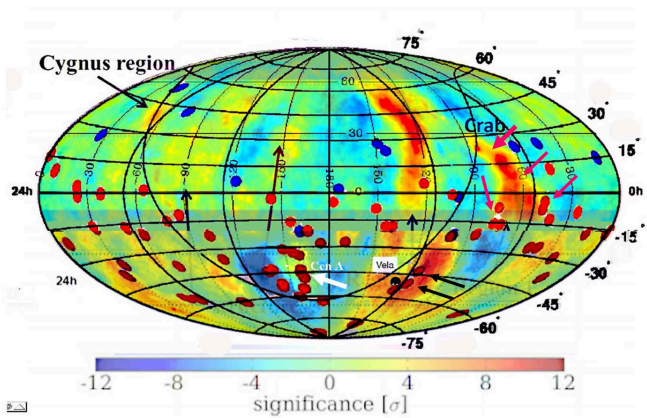}
  \caption{
  The AUGER 2010 UHECR (red) and Hires (blue) event map in celestial coordinate on recent (2010) TeV diffused CR map (ARGO-Milagro-ICECUBE) and labels. See \cite{ARGO},\cite{Desiati}, and references. The triplet UHECR event clustering toward Vela and the TeV spread activity around  is remarkable. The Cen A clustering at the fig. center is the main feature in the UHECR map. ARGO TeV anisotropy born around Crab connect and overlap the UHECR events in nearby Orion TeV region. Note doublet along the galactic plane and the  TeV near Cygnus and Cas A regions, where Hires did and we foresee TA may detect UHECR events.}
  \label{fig4-5}
 \end{figure}


\section{Conclusions: UHECR-TeV by $Ni^{57}$ decay? }
  Surprisingly the UHECR puzzle maybe at a corner stone: the UHECR-Multiplet along Cen A, the absence of Virgo, the hint of correlation with  Vela and  with galactic TeV anisotropy \cite{Desiati} \cite{ARGO}, might be in part solved by an extragalactic lightest nuclei, mainly He, from Cen A , (see Fig.\ref{fig0},\ref{fig1}, \ref{fig2}) \cite{Fargion2008}; their composition  is coexisting with present slant depth data (See Fig.\ref{fig3},\ref{fig4}), cite\cite{Auger10}. A partial confirm is the predicted \cite{Fargion2011} and observed \cite{Auger11} multiplet clustering by fragments (D,p) at half UHECR edge energy aligned   with Cen A: He like UHECR  maybe bent by a characteristic angle as large as  $\delta_{rm-He}  \simeq 16^\circ$ \cite{Fargion2011};  fragments multiplet along tails spread  at $\delta_{rm-p}  \simeq 32^\circ$ \cite{Fargion2011},\cite{Auger11}.  Other UHECR spread events, might be due to a  dominant heavy radioactive nuclei component  $Ni^{56}$, $Ni^{57}$ (and $Co^{56}$, $Co^{60}$,originated  by galactic sources  (old SNR-GRB relics, see Fig.\ref{fig4-5}) as suggested also by relic $Al$ nuclei at rest in gamma map  (see Fig.\ref{fig5}). UHECR Ni,Co maybe deflected by $18,7^{o}$ for Vela,  $128^{o}$ (or less) for Crab tuning within TeV inhomogeneities, made by boosted hundred keV gamma and beta positrons decay, shining at TeVs .  We predict here analogous UHECR traces around Cygnus and Cas A in future TA UHECR  maps. Inner galactic core UHECR are widely spread and hidden by magnetic fields in dense magnetic galactic core arms. However more clustering around ($\geq 20^{o}$) the galactic plane far from the core, is expected in future data. Magellanic cloud and stream may rise in UHECR maps. UHECR should rise around Cas A and Cygnus, seen by T.A. in North sky.
   The UHECR spectra cut off maybe not indebt to the expected extragalactic GZK feature but to the more modest imprint of a galactic confinement and of nuclei spectrography.    The UHECR radioactive beta decay in flight may trace in new $\nu_{\tau}$ neutrino astronomy or anisotropy, noise free,  related to astronomical (parasite oscillated) tau neutrino;   boosted tau (\emph{mini-double bang}  within a 5 meter size) in Deep Core  or Antares \cite{antares} may reveal hundred TeV tau decay (seeing  similar PeV ones in ICECUBE \cite{Learned}). Also Tau airshowers may rise in Cherenkov beamed air-showers. \cite{Fargion1999}, \cite{FarTau},\cite{Aita11} or in fluorescence telescopes at higher energies \cite{FarTau},\cite{Feng02},\cite{Bertou2002},\cite{Auger07}, \cite{Auger08}.  The discover of such expected  Neutrino astronomy  may shed additional light on the UHECR nature, origination  and mass composition, while  opening our eyes to mysterious  UHECR sources. Future gamma data and UHECR correlation, additional multiplets  may  lead to a more conclusive fit of this unsolved, century old, cosmic ray puzzle. Tau astronomy (observable also by mini-twin double bangs in Deep Core or Tau Airshowers) may also offer an additional test and an independent  hint on UHECR nature.\\

\clearpage

\end{document}